\documentclass[sigconf]{acmart}
\usepackage[utf8]{inputenc}
\AtBeginDocument{%
  \providecommand\BibTeX{{%
    \normalfont B\kern-0.5em{\scshape i\kern-0.25em b}\kern-0.8em\TeX}}}

\copyrightyear{2023} 
\acmYear{2023} 
\acmConference[ACM SIGMOD]{ACM International Conference on the Management of Data}{June 18--23,
  2023}{Seattle, WA}
\acmBooktitle{Companion of the 2023 International Conference on Management of Data (SIGMOD-Companion '23), June 18--23, 2023, Seattle, WA, USA}
\setcopyright{acmlicensed}\acmConference[SIGMOD-Companion '23]{Companion of the 2023 International Conference on Management of Data}{June 18--23, 2023}{Seattle, WA, USA}
\acmBooktitle{Companion of the 2023 International Conference on Management of Data (SIGMOD-Companion '23), June 18--23, 2023, Seattle, WA, USA}
\acmPrice{15.00}
\acmDOI{10.1145/3555041.3589410}
\acmISBN{978-1-4503-9507-6/23/06}

\usepackage{balance}

\title{Data Processing with FPGAs on Modern Architectures}

\author{Wenqi Jiang}
\affiliation{%
Systems Group, \\Department of Computer Science, \\ ETH Zürich
  \country{Switzerland}
}
\email{ wenqi.jiang@inf.ethz.ch }

\author{Dario Korolija}
\affiliation{%
Systems Group, \\Department of Computer Science, \\ ETH Zürich
  \country{Switzerland}
}
\email{ dario.korolija@inf.ethz.ch }

\author{Gustavo Alonso}
\affiliation{%
Systems Group, \\Department of Computer Science, \\ ETH Zürich
  \country{Switzerland}
}
\email{ alonso@inf.ethz.ch }

\begin{document}

\begin{abstract}
Trends in hardware, the prevalence of the cloud, and the rise of highly demanding applications have ushered an era of specialization that is quickly changing the way data is processed at scale. These changes are likely to continue and accelerate in the next years as new technologies are adopted and deployed: smart NICs, smart storage, smart memory, disaggregated storage, disaggregated memory, specialized accelerators (GPUS, TPUs, FPGAs), as well as a wealth of ASICs specifically created to deal with computationally expensive tasks (e.g., cryptography or compression). In this tutorial, we focus on data processing on FPGAs, a technology that has received less attention than, e.g., TPUs or GPUs but that is, however, increasingly being deployed in the cloud for data processing tasks due to the architectural flexibility of FPGAs, along with their ability to process data at line rate, something not possible with other types of processors or accelerators. 

In the tutorial, we will cover what FPGAs are, their characteristics, their advantages and disadvantages over other design options, as well as examples from deployments in the industry and how they are used in a variety of data processing tasks. Then we will provide a brief introduction to FPGA programming with High-Level Synthesis (HLS) tools as well as briefly describe resources available to researchers in the form of academic clusters and open-source systems that simplify the first steps. The tutorial will also include several case studies borrowed from research done in collaboration with companies that illustrate both the potential of FPGAs in data processing but also how software and hardware architectures are evolving to take advantage of the possibilities offered by FPGAs. These use cases include: (1) Approximated Nearest Neighbor Search (ANNS), to illustrate the problem of searching and processing large vector data collections, a problem relevant in both traditional data management and machine learning, (2) remote disaggregated memory, showing how the cloud architecture is evolving and demonstrating the potential for operator offloading and line rate data processing, and (3) recommendation systems which stand in for applications with very tight latency constraints that must, nevertheless, process vast amounts of data to provide results of sufficiently high quality. 
\end{abstract}

\ccsdesc[500]{Computer systems organization~Parallel architectures}
\ccsdesc[500]{Information systems~Data management systems}

\keywords{Data management, data processing, hardware acceleration, FPGA}

\maketitle

\section{Introduction}

Three important trends are significantly changing how data is processed at scale and, consequently, affecting the design of data processing systems and databases, as well as opening many opportunities for research at all levels of the architecture. These trends can be summarized as follows. 

First, CPUs are no longer the main or preferred computing processors when processing data. This is because general-purpose CPUs no longer provide the performance needed in many applications \cite{Thompson:2021:DCGP} and, as a result, data processing tasks are quickly being shifted towards GPUs, TPUs, or FPGAs. An example of the interesting implications of such a shift is projects such as Microsoft's Hummingbird \cite{KoutsoukosNKSAI21,Hummingbird22}, which takes advantage of the large investments being made in tensor computations by mapping data processing operations into tensor operations that can be run in the increasing amount of hardware and software available for tensor processing. 

Second, the cloud has become the dominant computing platform, removing many of the barriers that existed in the past to adopt highly specialized designs whose complexity can be managed and absorbed by the cloud provider without exposing it to the end user. These designs also help the cloud provider to make deployments more efficient by offloading system tasks to accelerators rather than using CPU cycles for that purpose \cite{Caulfield:2016:AccArch,Firestone:2018:SmartNICs}. The large economies of scale in the cloud provide the ideal background to deploy specialized designs that go away from CPU-based data processing and shift those tasks towards other parts of the overall system architecture, e.g., to caching layers with accelerators \cite{AWSaqua}.  

Third, there is an increasing number of highly demanding applications that require to process ever-growing amounts of data with very tight response time constraints and high accuracy requirements. Such strict Service Level Agreements (SLAs) can often not be met using conventional architectures due to the intrinsic inefficiencies of general-purpose computers, the data movements induced by large-scale distributed data processing, and the inability to process data close to the source which causes further inefficiencies by requiring to move data that is not needed but that is treated as a unit (a table, a segment, a file, a page or a block) by the underlying systems. Concrete examples involving data processing engines include Alibaba's X-Engine \cite{Huang:XEngine:2019}, which uses FPGAs to accelerate the management of Log Merge Trees in databases to ensure the latency constraints required by a large e-commerce site are met \cite{LSM20}, or Microsoft's KV-Direct \cite{KV-Direct} which uses an FPGA based smart NIC to accelerate access to Key-Value Stores through RDMA.

In the interesting and varied system architecture landscape that these three trends are shaping, the possibilities offered by FPGA-based design are of particular interest as demonstrated by their growing adoption in the industry for a wide range of data processing tasks \cite{FPGAs-Future,FPGAs-Clients}. FPGAs are not a replacement for CPUs, GPUs, or TPUs, but configure different parts of the design space offering advantages that other current options do not have: line rate processing \cite{Linerate20}, enabling processing streams of data out of the network, disks, or memory without performance loss; architectural flexibility in that they can be inserted in places where the other type of processor cannot be used \cite{AccordA} such as in NICs, in storage, in memory, etc. using \textit{bump-on-the-wire} designs that greatly facilitate their deployment; and the customizable nature of reconfigurable computing with the FPGA serving equally well to accelerate network function virtualization \cite{Firestone:2018:SmartNICs}, data reorganization in a database \cite{Huang:XEngine:2019}, or to accelerate joins \cite{Joins20}.

In this 3 hours tutorial, we will provide an in-depth hands-on lecture on FPGAs, their advantages and disadvantages, as well as cover use cases demonstrating their potential. We will also point participants to resources available to academics that have made it much easier to explore FPGA data processing without having to incur the expenses of building an FPGA-based infrastructure as well as to a wide range of open-source tools that significantly simplify the first steps by providing examples and basic software modules for data processing.

The presenters of this tutorial have extensive experience in FPGA designs and have successfully completed many projects both in an academic context but also in collaboration with industry with companies such as HPE \cite{Farview22}, Alibaba \cite{FleetRec21}, Amadeus \cite{DecisionTrees19,Maschi20}, SAP \cite{Chiosa22}, or AMD/Xilinx \cite{HePPOAB21}. The group runs one the AMD Heterogeneous Acceleration Computing Clusters (https://www.amd-haccs.io/) in a configuration specifically designed to support distributed data processing (https://systems.ethz.ch/research/data-processing-on-modern-hardware/hacc.html). The tutorial will explain how to gain access to such facilities, how to use them, and also point out useful tools available to software programmers to take the first steps in programming FPGAs. 

The intended audience is academic researchers with some experience in system-level programming and who are interested in exploring the opportunities offered by emerging hardware architectures. The tutorial will also be relevant for researchers who want to gain a wider perspective on how the cloud and hardware trends are affecting the underlying computing platforms and the impact that such changes will have on data processing. 

\section{Tutorial Contents}

The tutorial will strike a balance between motivating the technology, explaining how to use it, and providing use cases that demonstrate its potential. The use cases are based on our research as that way we can provide insights on the designs, give access to the open source code implementing the designs, and perform hands-on demonstrations. These use cases have been selected to cover a wide range of potential applications in data processing, machine learning, operators, systems support, as well as engine architecture, thereby covering an ample spectrum of potential research interests. We will also explain how to use a cluster of FPGAs, still rarely used in research but increasingly relevant in the industry, to give the participants the opportunity to gain an edge on their research. 

The three hours tutorial will be organized as follows: 

\begin{itemize}

\item{\textbf{Introduction and Background}} We will start the tutorial with a brief overview of the trends explained above and numerous examples from the industry on how FPGAs are being used to solve complex system design and demanding performance problems not only in data processing but also in other related areas such as search or machine learning. We will also cover what FPGAs are, without getting into architectural details but emphasizing their main characteristics and also explaining how they are evolving to more sophisticated platforms that it is typically acknowledged. We will discuss their differences with respect to GPUs and CPUs and provide concrete use cases where they are being successfully used to implement data processing operations that are otherwise expensive, emphasizing why FPGAs are particularly useful for those purposes and why other alternatives are not as optimal. This section will help participants less familiar with the technology to gain an understanding of the technology and why it is being deployed in practice. It will also help to motivate the next section in terms of how to program them and where they can be successfully employed. 

\item{\textbf{Programming}} This section will start by explaining the differences between programming in a temporal architecture (CPU, GPU) and a spatial architecture (FPGA) and what that means in practical terms when optimizing designs. We will also briefly discuss the implications in terms of general applicability in terms of threads/processes, multi-tenancy, shared resources, etc. We will then provide examples of programming simple tasks using High-level Synthesis (HLS) in C/C++ and the use of pragmas to achieve the required level of parallelism. Finally, we will discuss the different types of parallelism available in an FPGA, how they differ from those available in a CPU, and how they can be used to achieve greater performance in spite of the lower clock frequencies available in an FPGA. The duration of the tutorial does not allow us to give a full-fledged coverage of programming in HLS but the time devoted to the topic should be enough for participants with previous programming experience to get a solid idea of how it looks like and to motivate them to try simple designs. 

\begin{figure}[t]
  \centering
  \includegraphics[width=0.92\linewidth]{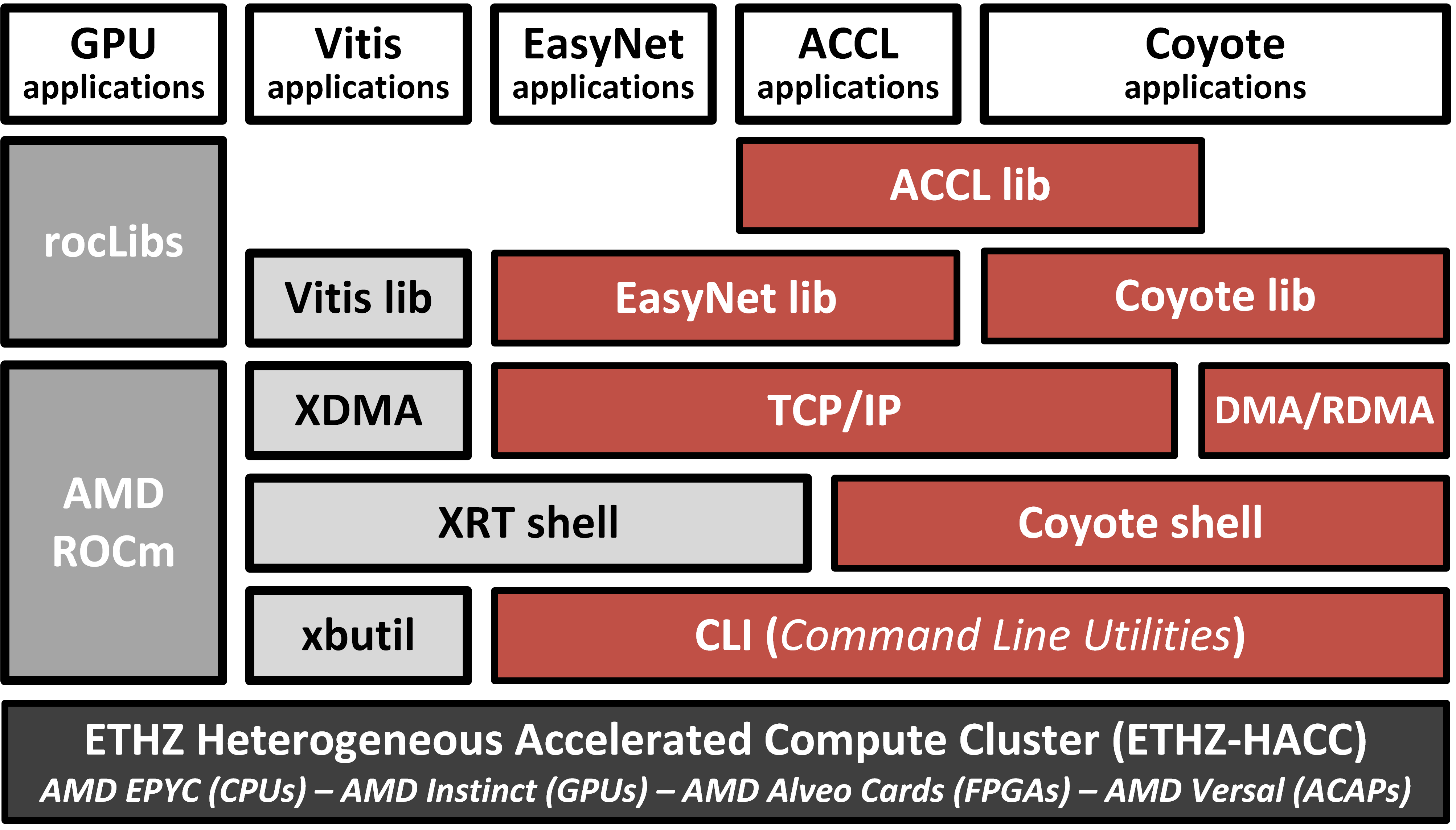}
  \vspace{-0.7em}
  \caption{\textit{ETHZ HACC} - Hardware acceleration platform.}
   \vspace*{-3mm}
  \label{fig:hacc}
\end{figure}

\item{\textbf{Resources Available}} One of the barriers to doing research on data processing using FPGAs is access to FPGAs, especially at the scale needed to run distributed applications. Another limitation is the lack of availability of good open-source examples and supporting infrastructure to build more complete systems. Fortunately, this situation is slowly starting to change, and in this section we will cover both access to public computing resources available to academia for research purposes, shown in Figure~\ref{fig:hacc}, as well as a number of open-source tools that make it much easier to explore advanced designs. These tools and systems include support for networking (e.g., TCP/IP \cite{HeKA21,RuizSSAL19}, RDMA \cite{SidlerWCKA20} and shells for multi-tenancy and easy integration in a data center \cite{KorolijaRA20}) as well as numerous coding examples of operators \cite{Chiosa22,KulkarniCPKSA20,HeWA20} in addition to complete systems implementing important data processing functionality \cite{FleetRec21,MicroRec21,DecisionTrees19,Maschi20}. Some of these coding examples will be covered in depth during the use cases. 

\item{\textbf{Use Case I: Smart disaggregated Memory with operator offloading}} As a first use case we have selected one based on conventional database architectures and the impact that cloud designs are having on them. The use case is based on Farview, shown in Figure~\ref{fig:farview}, a system developed together with HPE \cite{Farview22} implementing disaggregated DRAM memory attached to the network and capable of supporting the offloading of relational operators to the memory to reduce data movement and improve overall performance. From a design perspective, the use case is interesting as it illustrates several important aspects for database researchers: operator designs, line rate processing on data streams coming in or out of memory and the network, and how to use novel architectures to accommodate hardware developments. The design also incorporates an open-source RDMA stack that brings it to a competitive level with existing commercial solutions and shows the flexibility provided by FPGAs to break through the  system layers. The use case should be easily understandable to a database audience as the motivation, and the results are clearly explainable in terms of performance improvements to a traditional relational engine. The design is also related to that used by Amazon in Redshift to offload SQL operators to a caching layer \cite{AWSaqua}, thereby providing the audience with insights on how such innovative systems actually operate in practice. 

\begin{figure}[t]
  \centering
  \includegraphics[width=0.82\linewidth]{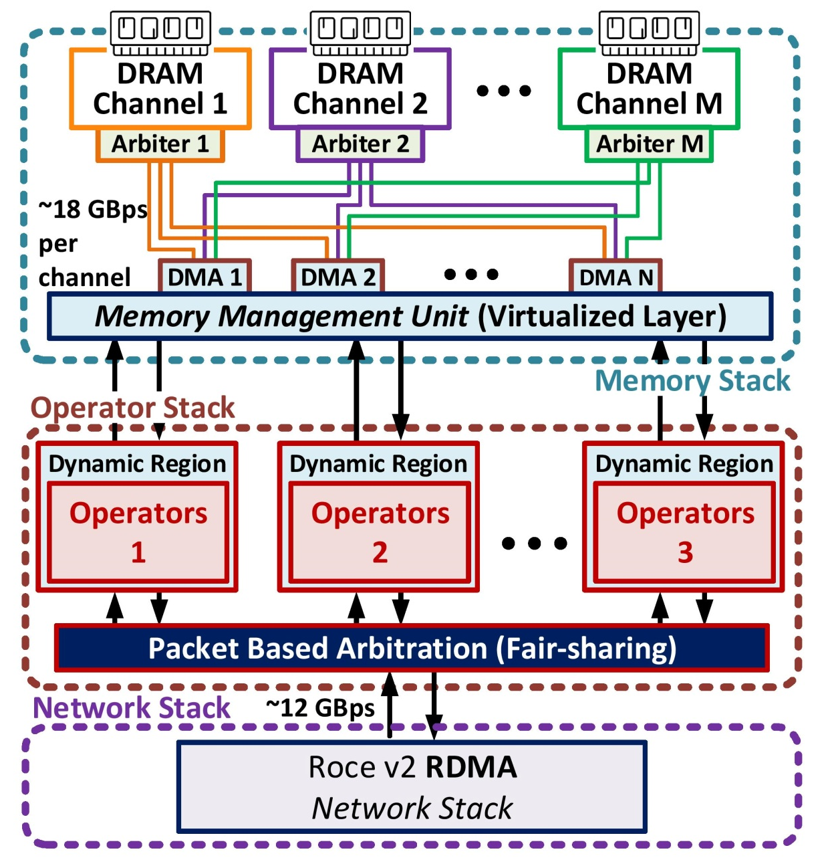}
  \vspace{-0.7em}
  \caption{High level view of Farview's architecture.}
   \vspace*{-3mm}
  \label{fig:farview}
\end{figure}

\item{\textbf{Use Case II: Approximated Nearest Neighbor Search}} The second use case bridges the gap between relational processing and ML/AI applications while still remaining in a  domain easily understandable to a database audience. Approximate nearest neighbor search (ANNS) is used to retrieve relevant data from large vector data collections~\cite{gao2023high}. Given a query vector, ANNS returns the K results that are similar to the query using some suitable distance metric. ANNS is now used extensively in ML applications, such as search engines\cite{chen2021spann, xiong2020approximate, karpukhin2020dense}, recommender systems~\cite{google_recommendation, suchal2010full}, and large language model training~\cite{guu2020realm, lewis2020retrieval, borgeaud2021improving}. These use cases also introduce strict SLAs that such designs face, which determine the type of trade-offs that can be done between accuracy (quality of the recall in this case) and performance goals. The use case is an extension of work we have done with industry on such ML systems (see next use case) and explores the problem of efficiently implementing ANNS in depth. ANNS is an algorithm extensively studied in the database community, so this section should be of interest to those who want to see the impact that specialized hardware has on the design and final performance of such algorithms. As before, the motivation and the nature of the results should be understandable to a database audience with a basic knowledge of NN algorithms, which will most likely be everybody in a conference like SIGMOD. The system we designed~\cite{jiang2023co} to accelerate ANNS is presented in Figure~\ref{fig:FANNS}.

\begin{figure*}[t]
  \centering
  \includegraphics[width=0.95\linewidth]{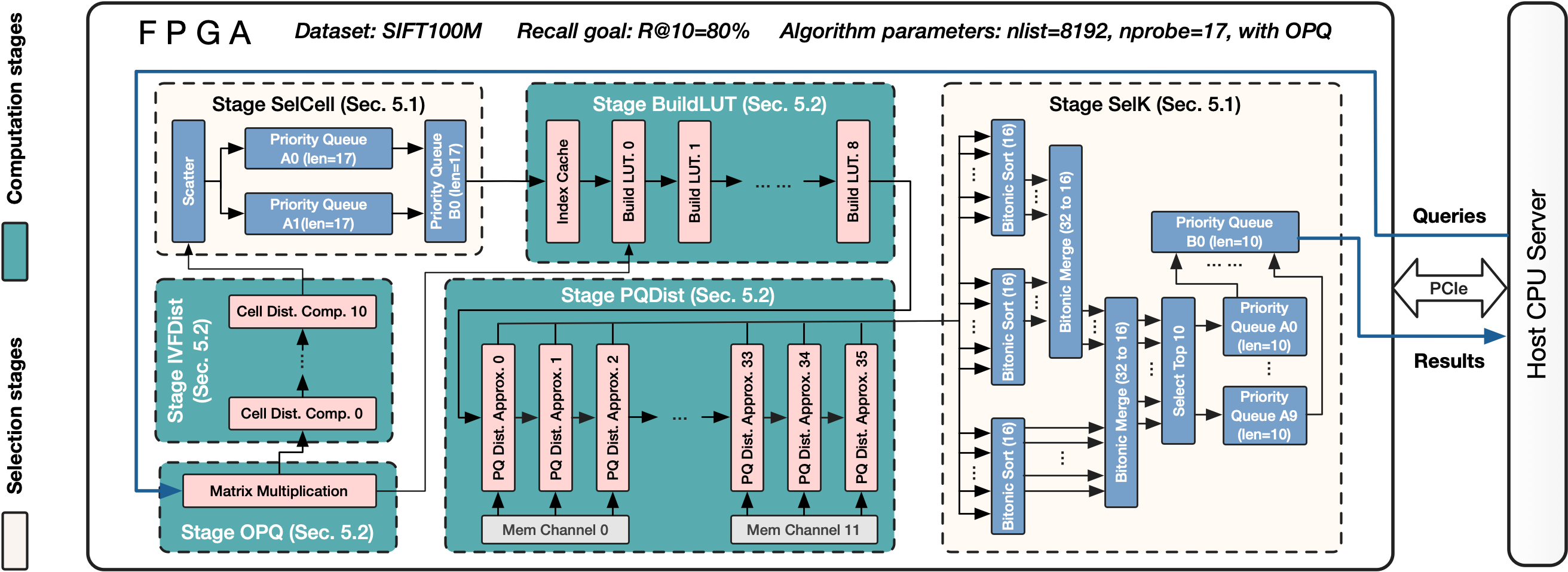}
  \vspace{-0.7em}
  \caption{FANNS: an approximate nearest neighbor search system accelerated by FPGA.}
  \label{fig:FANNS}
\end{figure*}

\item{\textbf{Use Case III: Recommendation Systems}} The third use case is based on work done with Aliababa on the recommendation systems used in their platform \cite{FleetRec21,MicroRec21}. The use case reaches out to those database researchers interested in machine learning applications, and it is a nice example of how to combine data processing techniques well known in the database community to make progress in difficult use cases in ML. As visualized in Figure~\ref{fig:rec_model}, a recommender model includes many embedding tables (typically tens to hundreds). An inference starts by looking up one embedding vector per table, before the fetched vectors are concatenated and fed into several fully-connected layers for the final click-through rate (CTR) prediction. Due to the many table lookups, the inference bottleneck shifts from computation (like regular DNNs) to memory accesses. To accelerate recommendation inference, we design the MicroRec accelerator that combines both data structure and hardware techniques to maximize inference performance. On the data structure side, we apply Cartesian products on a subset of tables to reduce the total number of memory accesses per inference. On the hardware side, we build an FPGA accelerator, as shown in Figure~\ref{fig:MicroRec}. The accelerator takes advantage of High Bandwidth Memory (HBM) available on FPGAs, such that we can allocate the tables to many banks. Smaller tables can be stored on FPGA's on-chip memory (SRAM) that can be accessed within a single clock cycle. The massive memory access parallelism of the accelerator results in one-order-of-magnitude speedup over CPUs. This design illustrates very well what can be done on an FPGA that cannot be done on other systems and why they achieve excellent performance in such use memory-bound use cases. 

\begin{figure}[t]
  \centering
  \includegraphics[width=0.95\linewidth]{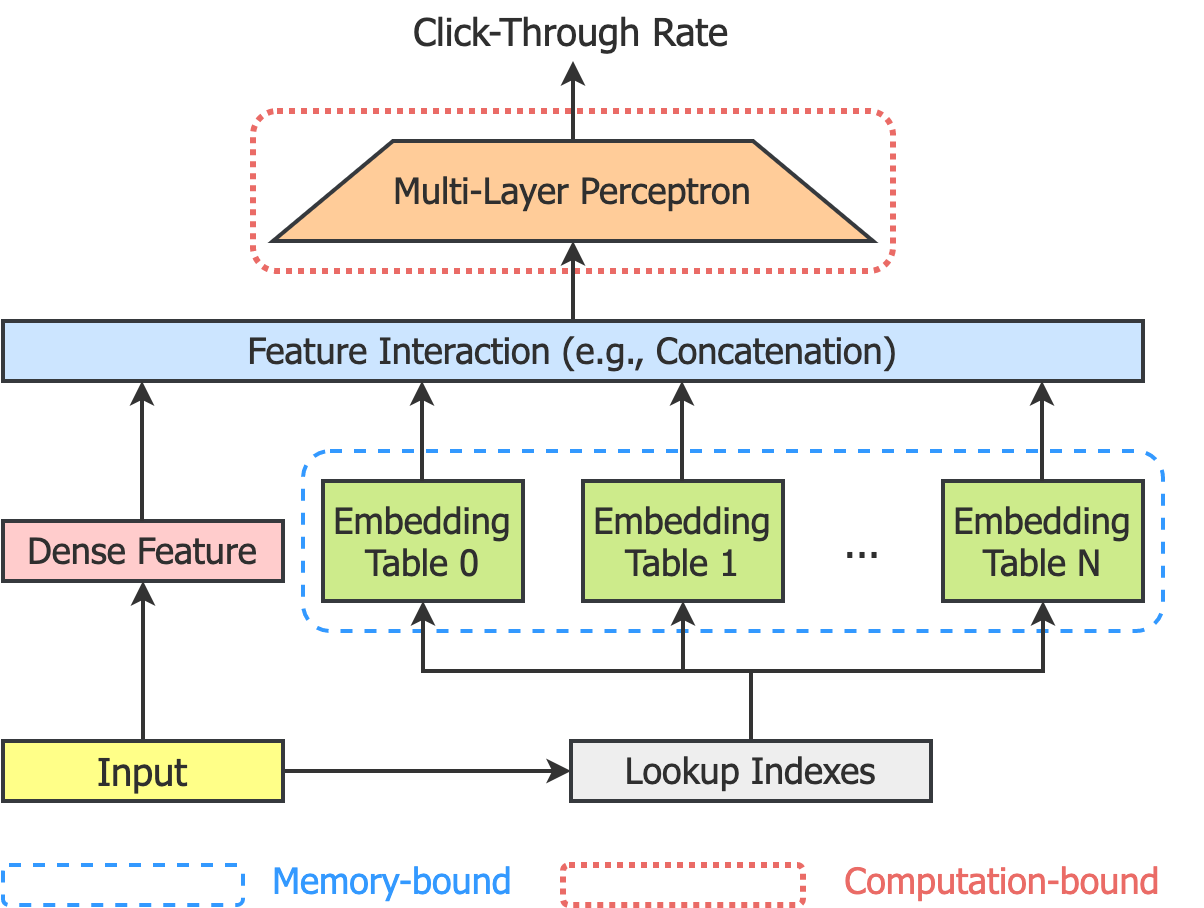}
  \vspace{-0.7em}
  \caption{A deep recommender model involves many table lookup operations during inference.}
  \label{fig:rec_model}
\end{figure}

\begin{figure}[t]
  \centering
  \includegraphics[width=0.95\linewidth]{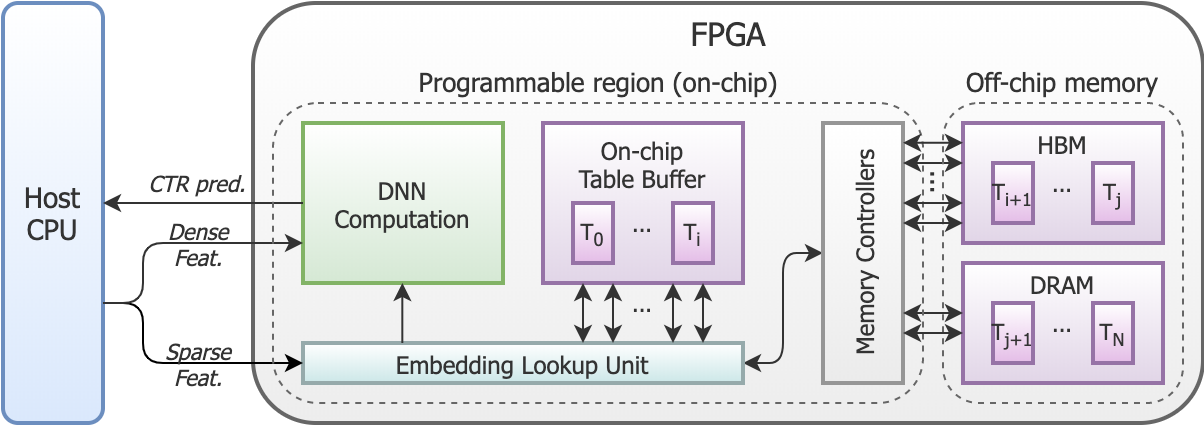}
  \vspace{-0.7em}
  \caption{\textit{MicroRec} is a high-performance recommendation inference engine on FPGA. It provides fast on-chip and off-chip feature retrieval and low-latency DNN computation simultaneously.}
  \label{fig:MicroRec}
\end{figure}

\item{\textbf{Use Case IV: Distributed Data Processing}} The final use case provides more of an overview of supporting infrastructure typically not available to researchers, but that is crucial to implement competitive designs and credibly tackle research problems involving larger data sets and cloud or data center deployments. The section will include systems like the recently published ACCL system \cite{HePPOAB21}, an MPI-like library that implements communication primitives for clusters of FPGAs to make it easier to implement operations such as \textit{all-reduce} or \textit{broadcast} that are the heart of many distributed data processing applications. We will also discuss memory management on FPGAs and point out existing systems that support large amounts of memory (up to 512 Gb) on the FPGA, at which point one can start thinking about radical designs like placing all the data of a medium-sized database in the memory of the FPGA instead of on the host memory. This section will not go into the details of the designs as those are less likely to be of interest to a database audience but it will provide the audience with the necessary material to use them in practice.

\item{\textbf{Discussion}} In the final part of the tutorial, we want to have an open discussion and question session where we can address concerns from the audience, discuss potential applications, provide hints of what research problems are most likely to benefit from using FPGAs, and encourage participants to explore this fascinating and timely topic. 

\end{itemize}

The tutorial will use mainly examples developed with tools, technology, and FPGA boards from AMD (Xilinx) and will encompass a number of software systems and hardware platforms: AMD Vitis, AMD HLS, as well as Xilinx UltraScale boards 280, 250, and U55c. All the use cases are based on open source designs already available to the community and that can serve both as a starting point to researchers eager to explore the area as well as baselines for those interested in improving the initial designs with new optimizations or expanding their scope to other use cases. 

\section{Intended Audience}

The tutorial has been designed with a database audience in mind and fully aware of the average level of knowledge available about reconfigurable computing. The presenters know both fields well, interact on a regular basis with both communities, and are in an ideal position to bridge the gaps between FPGA designers and database researchers. We expect the audience to be knowledgeable of database operators and database architecture. We also expect them to be familiar with basic cloud concepts such as storage disaggregation, resource management, and performance concerns at cloud level.  We will not assume previous knowledge of FPGAs or previous experience programming them. Nevertheless, those with such previous experience will still benefit from the in-depth analysis of the use cases and the discussion on what works and does not work in FPGA designs.

Given the time available, we have decided against having a hands-on tutorial with direct access to a cluster of FPGAs. While this would be very interesting, we consider it impractical given the expected audience as it would take too long to explain the setup, make sure everybody has a properly configured system in their laptops, and the basic usage of the tools is understood by everybody. Instead, we have opted for a tutorial that will empower those interested in taking the next step  by providing them with the information, examples, and support needed to start doing research on FPGA acceleration of data processing applications. 

\section{Presenters}

All presenters have experience teaching and presenting in front of large audiences. They also have extensive experience in FPGAs, data processing with FPGAs, and have developed numerous systems using the technology as part of academic research but also in close collaboration with industry. 

\textbf{Dario Korolija} is a final-year doctoral student at the Systems Group of the Department of Computer Science at ETH Zürich. He obtained his MSc degree from EPFL and completed his undergraduate studies at the University of Belgrade in Serbia.  Switzerland. He works at the intersection between software and hardware. His main research area is on creating novel abstractions for modern heterogeneous architectures working in the fields of computer architecture, data processing, operating systems and networking (mostly RDMA). He is also interested in recent compiler advancements (MLIR) and their usage for these novel computing systems. Dario has published at conferences such as CIDR, OSDI, ASPLOS, FPL, and FPGA. Web site: https://d-kor.github.io/

\textbf{Wenqi Jiang} is a third-year doctoral student at the Systems Group of the Department of Computer Science at ETH Zürich. He received his M.S. degree from Columbia University and B.Eng. from Huazhong University of Science and Technology, both with honors. His research interests include computer architecture, data management, and machine learning. More specifically, he is interested in enabling large-scale vector retrieval (approximate nearest neighbor search) by cross-stack solutions: from efficient retrieval algorithms to distributed systems design and high-performance hardware support. He has also explored hardware-accelerated solutions for industry-scale recommender systems in collaboration with Alibaba. Wenqi has published at conferences such as KDD, MLSys, and FPL.  Web page: https://wenqijiang.github.io/

\textbf{Gustavo Alonso} is a professor at the Systems Group of the Department of Computer Science at ETH Zürich. His research interests include data management, distributed systems, cloud computing architecture, and hardware acceleration through reconfigurable computing. Gustavo has served as PC chair for conferences in several areas including VLDB, ICDE, EDBT, EuroSys, Middleware, and ICDCS and regularly serves in the Program Committee of CIDR, VLDB, SIGMOD, FPGA, ATC, EuroSys,  OSDI, and MLSys. He was a member of the VLDB Endowment and the EDBT Executive Board and the Chair of EuroSys, the European Chapter of ACM SIGOPS. Gustavo has received 4 Test-of-Time Awards for his research in databases, software runtimes, middleware, and mobile computing. He is an ACM Fellow, an IEEE Fellow, and a Distinguished Alumnus of the Department of Computer Science of UC Santa Barbara. Web page: https://people.inf.ethz.ch/alonso/

\section{Summary}

This 3 hours tutorial will provide the audience with an opportunity to gain insights on an increasingly relevant technology (data processing on FPGAs) that is rapidly being deployed in the industry, especially by cloud providers, but that has not yet received all the attention from the academic research community it deserves. The tutorial is intended to serve as a demonstrator of the potential of the technology, a survey of recent  deployments and existing related work, a pointer to available resources, and an illustration that starting with FPGA programming is less daunting than it appears. As such, the tutorial can be of interest to both young researchers considering possible research directions as well as for established researchers interested in the impact that the new technology is having and will have on data processing. 

\bibliographystyle{ACM-Reference-Format}
\balance
\bibliography{References}

\end{document}